\documentclass[12pt]{iopart}
\usepackage{graphicx}

\begin{document}

\title{Electron spin resonance study of the LaIn$_{3-x}$Sn$_x$ superconducting system}

\author{E M Bittar$^{1,2}$, C Adriano$^1$, C Giles$^{1}$, C Rettori$^{1,3}$, Z Fisk$^{4}$ and P G Pagliuso$^{1,4}$}

\address{$^1$ Instituto de F\'{\i}sica ``Gleb Wataghin", UNICAMP, 13083-859, Campinas, SP, Brazil}
\address{$^2$ Laborat\'{o}rio Nacional de Luz S\'{\i}ncrotron, C. P. 6192, 13083-970, Campinas, SP, Brazil}
\address{$^3$ Centro de Ci\^{e}ncias Naturais e Humanas, Universidade Federal do ABC, 09210-170, Santo Andr\'{e}, SP, Brazil}
\address{$^4$ Department of Physics and Astronomy, University of California Irvine, 92697-4575, Irvine, CA, USA}

\ead{eduardo.bittar@lnls.br}

\begin{abstract}
The LaIn$_{3-x}$Sn$_x$ alloy system is composed of superconducting Pauli paramagnets. For LaIn$_3$ the superconducting critical temperature $T_c$ is approximately 0.7~K and it shows an oscillatory dependence as a function of Sn substitution, presenting its highest value $T_c$~$\approx$~6.4~K for the LaSn$_3$ end member. The superconducting state of these materials was characterized as being of the conventional type. We report our results for Gd$^{3+}$ electron spin resonance (ESR) measurements in the LaIn$_{3-x}$Sn$_x$ compounds as a function of $x$. We show that the effective exchange interaction parameter $J_{fs}$ between the Gd$^{3+}$ 4$f$ local moment and the $s$-like conduction electrons is almost unchanged by Sn substitution and observe microscopically that LaSn$_{3}$ is a conventional superconductor.
\end{abstract}

\section{Introduction}

The compounds of the LaIn$_{3-x}$Sn$_x$ alloy system are superconducting Pauli paramagnets. For LaIn$_3$ the superconducting critical temperature $T_c$ is approximately 0.7~K and it shows an oscillatory dependence as a function of Sn substitution, presenting its highest value $T_c$~$\approx$~6.4~K for the LaSn$_3$ end member~\cite{LaInSnTc,LaInSn1,LaInSn2}. This system was extensively studied between 1965 and 1975 partially due to the coexistence of superconductivity with a large temperature dependent magnetic susceptibility~\cite{LaInSn2}. This enhanced susceptibility was believed to result from a large exchange enhancement of the La $d$ band contribution. Also, the superconducting proprieties were attributed exclusively to the changes of the electronic states at the (In,Sn) site and the oscillatory behaviour of $T_c$ was credited to the nonmonotonic variation of the bare density of states as a function of Sn substitution~\cite{LaInSn2,SCAuCu3}. These materials were characterized as being conventional superconductors~\cite{LaSn3_Hc,LaSn3_gamma}. One evidence of this conventional superconductivity was seen by the suppression of $T_c$ as a function of magnetic impurities doping for LaSn$_3$~\cite{Tc_Gd_LaSn3}. The initial linear decrease of the critical temperature, in this case, followed the Abrikosov-Gorkov theory~\cite{AG} for a single gapped superconductor. Previously reported experimental parameters for this system are listed in Table~\ref{Tab_Exp_LaInSn1}.

\begin{table}
\caption{Previous reported experimental parameters for LaIn$_{3-x}$Sn$_x$. $\gamma$ is the Sommerfeld coefficient, $\lambda$ is the electron-phonon coupling parameter and $\Delta T_c/\Delta c$ is the suppression of $T_c$ by Gd$^{3+}$ magnetic impurities.} \label{Tab_Exp_LaInSn1} \centering
\begin{tabular}{llll}
\br
 &  $\gamma$ (mJ mol$^{-1}$ K$^{-2}$) & $\lambda^c$ & $\frac{\Delta T_c}{\Delta c}$ (K/\%)\\
\mr
LaIn$_{3}$ & 6.3$^{a}$ & 0.55(5) &  \\
LaIn$_{2.2}$Sn$_{0.8}$ &  & 0.65(5) &  \\
LaIn$_{1.5}$Sn$_{1.5}$ &  & 0.60(5) &  \\
LaSn$_{3}$ & 11.7$^{b}$ & 0.80(5) & -1.25(3)$^{d}$ \\
\br
\end{tabular}
\\
$^{a}$~Ref.~\cite{CeIn3_gamma};
$^{b}$~Ref.~\cite{LaSn3_gamma};
$^{c}$~Ref.~\cite{LaInSn2};
$^{d}$~Ref.~\cite{Tc_Gd_LaSn3}.
\end{table}

Our main goal is to study the LaIn$_{3-x}$Sn$_x$ alloys as a nonmagnetic analog for the CeIn$_{3-x}$Sn$_x$ heavy fermion system. Both series are isostructural and their crystal structures are cubic, space group $Pm$-$3m$ (n°.~221). For the Ce compounds the antiferromagnetic transition at $T_N$~=~10~K ($x$~=~0) was reported to decrease upon Sn substitution until it vanishes at a critical concentration of $x_c$~$\approx$~0.7~\cite{CeInSn_Lawrence,CeInSn_Euro}. In the vicinity of this point, logarithmic divergence of the electronic specific heat was seen, in accordance with the behaviour of heavy fermions near quantum critical instabilities~\cite{RevNFLStewart,RevLohneysen}. With further increase of Sn substitution ($x$~$\geq$~2.2) the Ce materials show valence fluctuation effects~\cite{CeInSn_Lawrence,CeInSn_Euro}. Here we report our results for Gd$^{3+}$ electron spin resonance (ESR) measurements in the LaIn$_{3-x}$Sn$_x$ compounds used as a reference for our ESR analysis of the CeIn$_{3-x}$Sn$_x$ series~\cite{CeInSn_ESR}. We extract the effective exchange interaction parameters between the Gd$^{3+}$ localized magnetic impurity and the conduction electrons in the normal state~\cite{ESR_SC} and compare it to the value obtained by previous study of the slope of the initial reduction of $T_c$ by the Gd$^{3+}$ impurities~\cite{Tc_Gd_LaSn3,AG}.

\section{Experiment details and results}

Flux-grown single crystals of Gd doped LaIn$_{3-x}$Sn$_x$ are synthesized~\cite{fiskandca}, where elemental La:Gd:In:Sn are weighted in the ratio 1-$y$:$y$:10-(10$x$/3):10$x$/3, with a nominal value for $y$ of 0.005 and $x$~=~0, 1.5 and 3. Polycrystalline samples are also grown by arc melting in Argon atmosphere. In this case the reactants ratio used is 1-$y$:$y$:3-$x$:$x$, with the same nominal value for $y$ and $x$~=~0, 0.7, 1.5 and 3. X-ray powder diffraction measurements confirm the cubic type structure for all compounds synthesized. The temperature dependence of the magnetic susceptibility $\chi(T)$ is measured for $2\leq T\leq300$~K, after zero field cooling. All ESR experiments are performed on a fine powder ($d\leq38$~$\mu$m) in a Bruker ELEXSYS $X$-band spectrometer (9.4~GHz) with a TE$_{102}$ cavity coupled to a helium-gas-flux $T$-controller system for $4.2\leq T\leq300$~K. Fine powder of crushed single crystals are used in the ESR experiments in order to increase the ESR signal-to-noise ratio. Through out this text, for convention, single crystalline samples are represented as closed symbols while polycrystals are seen as open symbols.

\begin{figure}
\begin{center}
\includegraphics[width=0.8\textwidth]{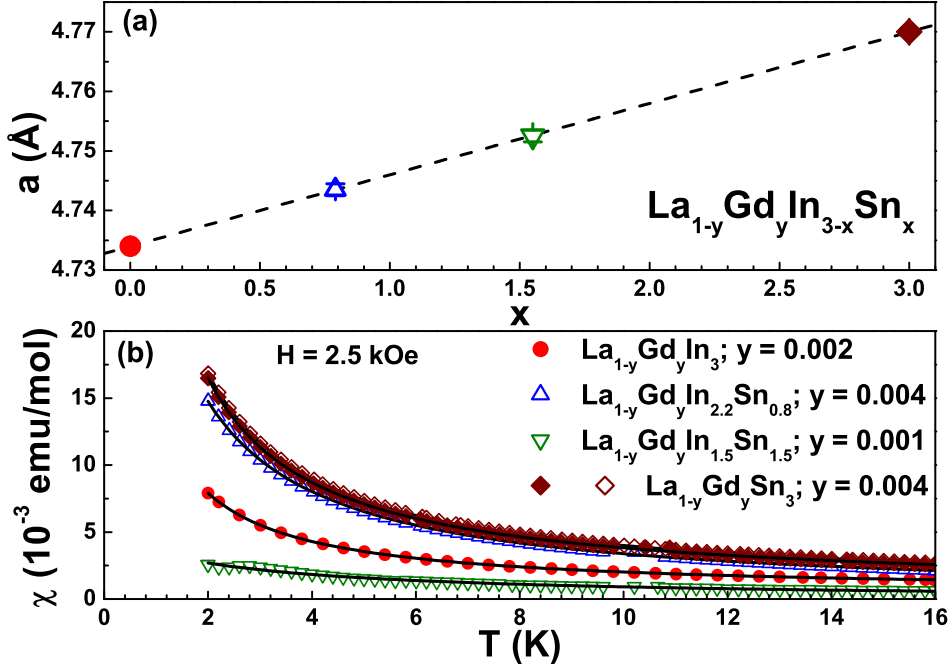}
\end{center}
\caption{Gd$^{3+}$ in La$_{1-y}$Gd$_y$In$_{3-x}$Sn$_x$: (a) cubic lattice parameter $a$ dependence as a function of $x$. The dashed line represents the Vegard law~\cite{CeInSn_Euro}. (b) Low temperature dependence of $\chi(T)$ at $H$~=~2.5~kOe. The solid lines are the Curie-Weiss fitting. The closed symbols identify the single crystalline samples and the open ones the polycrystals.}
\label{Fig1}
\end{figure}

To obtain the actual Sn concentration we use the Vegard's law, which states that the LaIn$_3$ cubic lattice parameter $a$ varies linearly with the doping of Sn~\cite{CeInSn_Euro}. Figure~\ref{Fig1}(a) illustrates the dependence of the cubic lattice parameter as a function of $x$, evidencing the synthesized samples. Figure~\ref{Fig1}(b) shows the temperature dependent magnetic susceptibility $\chi(T)$ data for La$_{1-y}$Gd$_y$In$_{3-x}$Sn$_x$. The curves are corrected for the core diamagnetism. From the Curie-Weiss law fitting of the low temperature data and knowing the effective magnetic moment of Gd$^{3+}$ ($\mu_{eff}$~=~7.94~$\mu_{B}$) we estimate the Gd doping concentration and its values are given in Table~\ref{Tab_Exp_LaInSn2}.

\begin{figure}
\begin{center}
\includegraphics[width=0.8\textwidth]{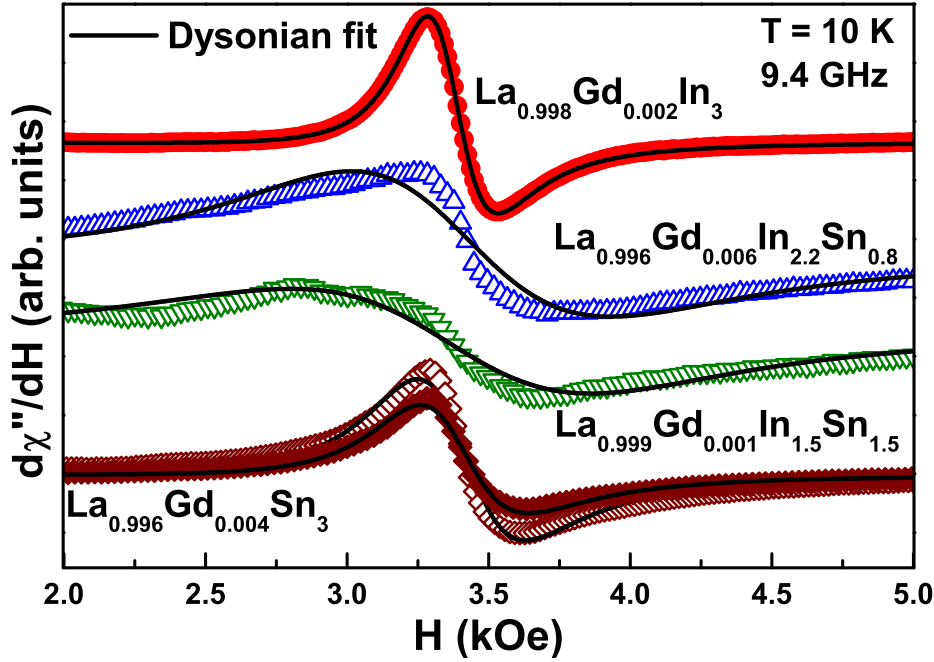}
\end{center}
\caption{$X$-band (9.4 GHz) Gd$^{3+}$ ESR powder spectra, at $T$~$\approx$~10~K, in La$_{1-y}$Gd$_y$In$_{3-x}$Sn$_x$. The solid lines are the Dysonian line shape analysis. The closed symbols identify the single crystalline samples and the open ones the polycrystals.}
\label{Fig_ESR}
\end{figure}

Figure~\ref{Fig_ESR} shows the Gd$^{3+}$ ESR $X$-band powder spectra in La$_{1-y}$Gd$_y$In$_{3-x}$Sn$_x$ at $T$~=~10~K. Each ESR spectrum consists of a single Dysonian resonance, characteristic of localized magnetic moments in a metallic host with a skin depth smaller than the size of the measured particles. From the fitting of the resonances to the appropriate admixture of absorption and dispersion Lorentzian derivatives, the $g$ values and linewidths $\Delta H$ are obtained~\cite{Feher}. The solid lines are the best fit to the observed resonances and the obtained $g$ values are presented in Table~\ref{Tab_Exp_LaInSn2}. Within the accuracy of the measurements, the $g$ values were temperature and Gd concentration independent (not shown). For La$_{0.996}$Gd$_{0.004}$In$_{2.2}$Sn$_{0.8}$ a background line is present in the spectrum at $H$~$\sim$~3.4~kOe and for La$_{0.999}$Gd$_{0.001}$In$_{1.5}$Sn$_{1.5}$ the ESR spectrum also show contribution of the background (1.5~$\leq$~$H$~$\leq$~2.5~kOe), since the signal in this sample is small due to low Gd$^{3+}$ concentration ($\approx$~0.1\%). For $x=3$ we observe that there is no appreciable difference in the resonance of grounded single and polycrystalline samples. This is not always the case in most systems and it cannot be established \textit{a priori} because polycrystals synthesized by arc-melting tend to present more inhomogeneity. Since in this work samples with different Sn content were grown by different techniques we show that there are no discernable differences in the ESR spectra for a sample grown by both processes.

\begin{figure}
\begin{center}
\includegraphics[width=0.8\textwidth]{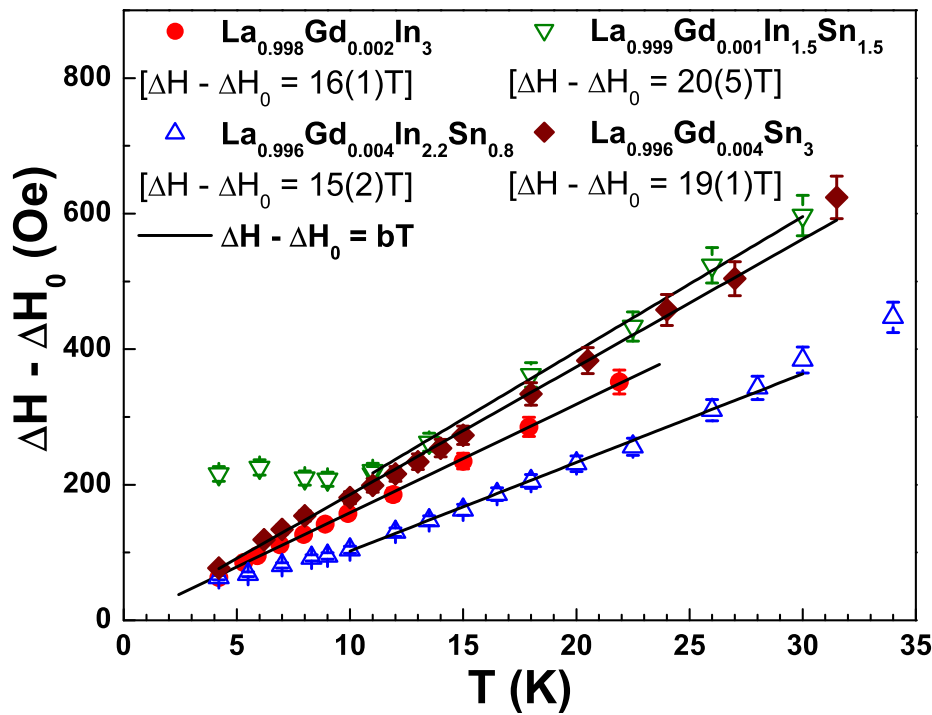}
\end{center}
\caption{Temperature dependence of the Gd$^{3+}$ ESR linewidth in La$_{1-y}$Gd$_y$In$_{3-x}$Sn$_x$. The solid lines are the best fit to $\Delta H$~-~$\Delta H_0$~=~$bT$. The closed symbols identify the single crystalline samples and the open ones the polycrystals.}
\label{Fig_Korringa}
\end{figure}

The temperature dependence of the Gd$^{3+}$ ESR linewidth in LaIn$_{3-x}$Sn$_x$ is seen in Figure~\ref{Fig_Korringa}. For all samples the width increases linearly with temperature. The linear dependence of the linewidth was fitted to the expression $\Delta H-\Delta H_0=bT$. The departure of the linear dependence decrease of $\Delta H$ at low temperature, seen for $x=1.5$ may be related to short range Gd-Gd interaction or an artifact of the fitting analysis because of the background contribution in this temperature range. The values for $\Delta H_0$ (residual linewidth) and $b$ (thermal broadening) are presented in Table~\ref{Tab_Exp_LaInSn2}.  The residual linewidth is defined as the ESR linewidth at $T=0$ and it follows the residual electrical resistivity $\rho_0$ behaviour, since both are dependent of the disorder. Its relatively large values for $0<x<3$ are mainly due to the disorder effects introduced by the In-Sn substitution (see Fig.~4 of Ref.~\cite{LaInSn1}). Within the accuracy of the measurements, the $g$ and $b$ values are Gd concentration independent for $y<2.0\%$ (not shown). Therefore, bottleneck and dynamic effects can be disregarded~\cite{ref20}.

\begin{table}
\caption{Experimental parameters for Gd$^{3+}$ diluted in La$_{1-y}$Gd$_y$In$_{3-x}$Sn$_x$.} \label{Tab_Exp_LaInSn2} \centering
\begin{tabular}{llllll}
\br
Gd$^{3+}$ in & $g$ & $\Delta g$ & $\Delta H_0$ (Oe)& $b$ (Oe/K)& $y$ \\
\mr
LaIn$_{3}$ & 2.020(5) & +0.027(5) & 50(5) & 16(1) & 0.002 \\
LaIn$_{2.2}$Sn$_{0.8}$ & 2.010(10) & +0.020(10) & 610(5) & 15(2) & 0.004  \\
LaIn$_{1.5}$Sn$_{1.5}$ & 2.020(10) & +0.030(10) & 470(10) & 20(5) & 0.001 \\
LaSn$_{3}$ & 2.010(10) & +0.020(10) & 140(5) & 19(1) & 0.004 \\
\br
\end{tabular}
\end{table}

\section{Analysis and discussion}

In the simplest treatment of the exchange interaction, $J_{fs}(q)\mathbf{S}\cdot\mathbf{s}$, between a localized 4$f$ electron spin ($\mathbf{S}$) on a solute atom (Gd$^{3+}$) and the free conduction electron spin ($\mathbf{s}$) of the host metal, the ESR $g$ shift $\Delta g$ (Knight shift)~\cite{Yosida} and the thermal broadening of the linewidth (Korringa rate)~\cite{Korringa}, when bottleneck and dynamic effects are not present~\cite{ref20}, can be written as:
\begin{equation}
\Delta g=J_{fs}\eta _{F}, \label{Eq_Deltag}
\end{equation}
and
\begin{equation}
b=\frac{d(\Delta H)}{dT}=\frac{\pi k_{B}}{g\mu _{B}}J_{fs}^{2}\eta _{F}^{2},  \label{Eq_Korringa}
\end{equation}
where $J$$_{fs}$ is the effective exchange interaction parameter between the Gd$^{3+}$ 4$f$ local moment and the $s$-like conduction electron in the absence of conduction electron momentum transfer~\cite{DavidovSSC12}, $\eta(F)$ is the ``bare" density of states for one spin direction at the Fermi surface, $k_B$ the Boltzman constant, $\mu_B$ the Bohr magneton and $g$ the Gd$^{3+}$ $g$ value in insulators.

For diluted rare-earths magnetic moments in intermetallic compounds with appreciable residual resistivity, i.e., large conduction electron spin-flip scattering, equations \ref{Eq_Deltag} and \ref{Eq_Korringa} are generally used in the analysis of the ESR data. In these cases, the equations above may be rewritten as
\begin{equation}
b=\frac{\pi k_{B}}{g\mu _{B}}(\Delta g)^{2}.
\label{eq3}
\end{equation}

\subsection{Analysis of the Gd$^{3+}$ ESR in LaIn$_{3}$}

The $g$ value of Gd$^{3+}$ ESR in a number of cubic insulating materials can be well averaged and approximated to 1.993(1) and has been established as the correct one for the ESR community for the last two decades~\cite{Abragam}. From the difference between the measured value (see table \ref{Tab_Exp_LaInSn2}) and $g=1.993(1)$, we calculate the $g$ shift $\Delta g$~$\simeq$~27(5)$\times 10^{-3}$ for the Gd$^{3+}$ resonance in La$_{0.998}$Gd$_{0.002}$In$_{3}$. Knowing that $\pi k_{B}/g\mu _{B}$~=~2.34$\times 10^{4}$~Oe/K, Eq.~\ref{eq3} predicts a thermal broadening of $b$~$\approx$~17(5)~Oe/K. This value is very close to the measured one, $b$~=~16(1)~Oe/K. Therefore, we can conclude that the approximations made in Eqs.~\ref{Eq_Deltag} and \ref{Eq_Korringa} are adequate for this compound and conduction electron-electron correlations~\cite{Moriya,Narath}, $q$ dependence of the exchange interaction~\cite{DavidovSSC12} and multiple band effects~\cite{YbAl3} do not need to be taken into account in the analysis of our ESR data. Indeed, nuclear magnetic resonance experiments showed that for LaT$_3$ (T~=~Sn, In, Pb, Tl, etc.)~\cite{LaInSnTc,LaInSn2,NMRLaSn3,NMRLaX3} conduction electron-electron correlations have negligible effect in these compounds. In the free conduction electron gas model, the Sommerfeld coefficient for a superconductor is given by $\gamma =(2/3)\pi ^{2}k_{B}^{2}\eta _{F}(1-\lambda)$, where $\lambda$ is the electron-phonon coupling parameter. For LaIn$_{3}$ $\gamma$~=~6.3~mJ/(mol~K$^2$)~\cite{CeIn3_gamma} and $\lambda$~=~0.55(5)~\cite{LaInSn2}. Hence, we estimate a total density of states at the Fermi level of $\eta _{F}$~=~0.8(1)~states/(eV~mol~spin). With the experimental and derived parameters above we can use Eq.~\ref{Eq_Korringa} to obtain the exchange interaction parameter between the Gd$^{3+}$ 4$f$ and a single $s$-like conduction band, $J_{fs}$~=~33(5)~meV, in LaIn$_{3}$.

\subsection{Analysis of the Gd$^{3+}$ ESR in LaIn$_{2.2}$Sn$_{0.8}$}

In La$_{0.996}$Gd$_{0.004}$In$_{2.2}$Sn$_{0.8}$, $\Delta g$~$\simeq$~20(10)$\times 10^{-3}$. Eq.~\ref{eq3} estimates $b$~$\approx$~10(10)~Oe/K, which is within experimental error is close to $b$~=~15(2)~Oe/K, found experimentally. Thus, the model of a single $s$-like band and no $q$ dependence can be used in the analysis of resonance in this material, as in the case of the Gd$^{3+}$ ESR in LaIn$_{3}$ seen above. Knowing that the density of states for LaIn$_{2.2}$Sn$_{0.8}$ is $\eta _{F}$~=~0.8(1)~states/(eV~mol~spin)~\cite{LaInSn2}, from Eq.~\ref{Eq_Korringa} we find $J_{fs}$~=~32(6)~meV (see Table~\ref{Tab_Cal_LaInSn}).

\subsection{Analysis of the Gd$^{3+}$ ESR in LaIn$_{1.5}$Sn$_{1.5}$}

The density of states of the LaIn$_{1.5}$Sn$_{1.5}$ sample is $\eta _{F}$~=~0.9(1)~states/(eV~mol~spin)~\cite{LaInSn2}. The $\Delta g$~$\simeq$~30(10)$\times 10^{-3}$ value observed relates to a Korringa rate $b$~=~20(15)~Oe/K by Eq.~\ref{eq3}, which is very close to the thermal broadening seen experimentally, $b$~=~20(5)~Oe/K. Consequentially, an analysis considering a single $s$-like band is valid as in both cases described above. From Eq.~\ref{Eq_Korringa} we obtain $J_{fs}$~=~32(8)~meV.

\subsection{Analysis of the Gd$^{3+}$ ESR in LaSn$_{3}$}

From the Sommerfeld coefficient expression and knowing that the LaSn$_{3}$ compound has $\gamma$~=~11.7~mJ/(mol~K$^2$)~\cite{LaSn3_gamma} and $\lambda$~=~0.80(5)~\cite{LaInSn2}, we estimate $\eta_{F}$~=~1.4(1)~states/(eV~mol~spin). The Korringa rate and $\Delta g$ found experimentally for La$_{0.996}$Gd$_{0.004}$Sn$_{3}$ are 19(1)~Oe/K and 20(10)$\times 10^{-3}$, respectively. From Eq.~\ref{eq3} we get $b$~$\approx$~10(10)~Oe/K, which indicates that, within experimental error, we can neglect multiple band and $q$ dependence effects. Therefore, we calculate, using Eq.~\ref{Eq_Korringa}, $J_{fs}$~=~20(2)~meV (see Table~\ref{Tab_Cal_LaInSn}). It is interesting to compare the exchange interaction parameter $J_{fs}$ obtained in the ESR experiment with the one obtained by the suppression of the superconducting transition temperature $T_c$ by Gd$^{3+}$ magnetic impurities $\Delta T_c/\Delta c$. From the Abrikosov-Gorkov theory~\cite{AG} we have:
\begin{equation}
\left|\frac{\Delta T_c}{\Delta c}\right|=\frac{\pi^2}{8k_B}J_{fs}^2\eta_{F}(g_J-1)^2J(J+1),  \label{Eq_AG}
\end{equation}
where $g_J$ is the Land\'{e} factor. The value of $\Delta T_c/\Delta c=-1.25(3)$~K/\% reported in Ref.~\cite{Tc_Gd_LaSn3} when used in Eq.~\ref{Eq_AG} predicts $J_{fs}$~$\approx$~20~meV, in accordance with the one obtained by ESR. This result provides microscopical evidence of the conventional character of the superconducting state in LaSn$_{3}$ and also corroborates that the mechanism of relaxation of the resonance of the Gd$^{3+}$ ion in this compound is via a single $s$-like conduction band~\cite{ESR_SC}.

\begin{table}
\caption{Derived parameters for Gd$^{3+}$ diluted in LaIn$_{3-x}$Sn$_x$.} \label{Tab_Cal_LaInSn} \centering
\begin{tabular}{llll}
\br
 &  & ESR & SC \\
Gd$^{3+}$ in & $\eta_F$ (states eV$^{-1}$ mol$^{-1}$ spin$^{-1}$) & $J_{fs}$ (meV) & $J_{fs}$ (meV) \\
\mr
LaIn$_{3}$ & 0.8(1) & 33(5) & \\
LaIn$_{2.2}$Sn$_{0.8}$ & 0.8(1)$^a$ & 32(6) & \\
LaIn$_{1.5}$Sn$_{1.5}$ & 0.9(1)$^a$ & 32(8) & \\
LaSn$_{3}$ & 1.4(1) & 20(2) & $\approx$20 \\
\br
$^{a}$~Ref.~\cite{LaInSn2}
\end{tabular}
\end{table}

\begin{figure}
\begin{center}
\includegraphics[width=0.8\textwidth]{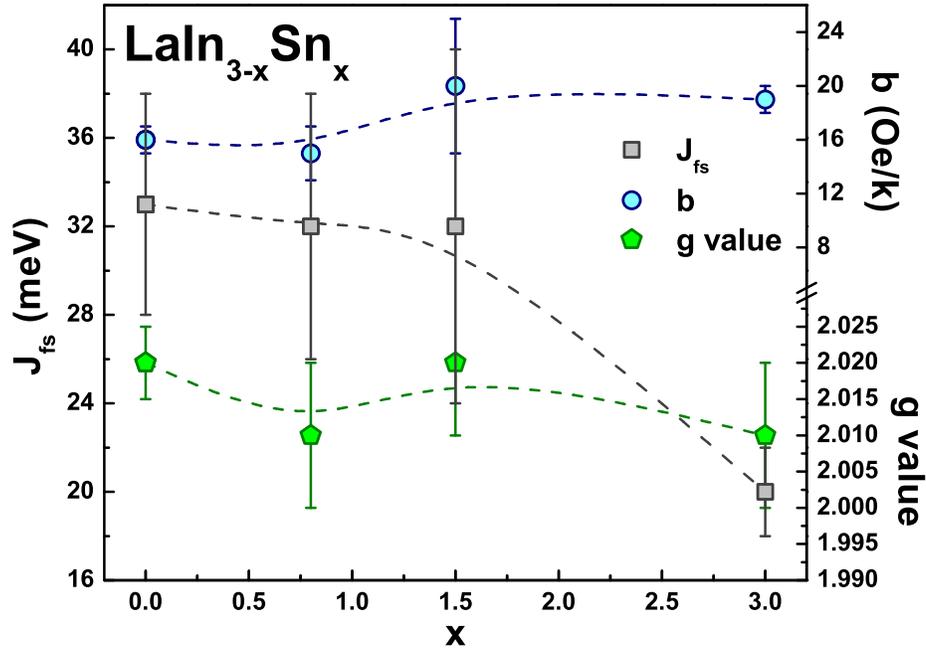}
\caption{Evolution of the effective exchange interaction parameter $J_{fs}$, Korringa rate $b$ and $g$ value of the Gd$^{3+}$ ESR in LaIn$_{3-x}$Sn$_x$ as a function of $x$. The dashed lines are guides to the eyes.}
\label{Fig_CeInSn_Conj}
\end{center}
\end{figure}

Figure~\ref{Fig_CeInSn_Conj} summarizes some experimental and derived parameters, obtained for the Gd$^{3+}$ ESR in the LaIn$_{3-x}$Sn$_x$ system. We observe that the $g$ shift and the Korringa rate do not change significantly with the Sn substitution. From the analysis of the experimental data we showed that the Gd$^{3+}$ resonance relaxation in these compounds occurs via a single $s$-like conduction band that is independent of momentum transfer. We also obtain the values of the effective exchange interaction parameter $J_{fs}$, which tends to decrease with the substitution of In by Sn. Since the LaIn$_{3-x}$Sn$_x$ compounds increase the cubic lattice parameter (Fig.~\ref{Fig1}) as a function of $x$ this may be responsible for decreasing $J_{fs}$.

\section{Summary}

In summary, we report Gd$^{3+}$ ESR results in the LaIn$_{3-x}$Sn$_x$ superconducting system, which permitted the extraction of the effective exchange interaction parameter $J_{fs}$ between the Gd$^{3+}$ 4$f$ local moment and the $s$-like conduction electrons in these compounds. We showed that $J_{fs}$ slightly decreases as a function of Sn substitution. This is in sharp contrast to the results found for the Gd$^{3+}$ resonance in the CeIn$_{3-x}$Sn$_x$ heavy fermion analog system, much due to the peculiarities of the magnetic compounds~\cite{CeInSn_ESR}. We have also observed microscopically that the superconductivity is of the conventional type in LaSn$_{3}$, i.e., the exchange coupling between the Gd$^{3+}$ local moment and the conduction electrons governs the impurity relaxation and pair-braking processes.

\section*{Acknowledgments}

We would like to thank J. C. B. Monteiro and F. C. G. Gandra for helping with the polycrystalline samples synthesis. This work was supported by FAPESP, CNPq, FINEP and CAPES (Brazil).

\end{document}